\documentclass{article}

\usepackage{microtype}
\usepackage{graphicx}
\usepackage{subfigure}
\usepackage{booktabs} 

\usepackage[font=small,skip=0pt]{caption}
\usepackage{enumitem}
\setlist{nolistsep}

\usepackage[hyphens]{url}
\usepackage{hyperref}

\usepackage{yk}
\usepackage{tikz}
  \usetikzlibrary{bayesnet}

\def\db{{\sf db}}
\def\hA{\widehat{A}}

\def\MN{{\sf MN}}

\usepackage[accepted]{icml2018}

\icmltitlerunning{Debiasing representations by removing unwanted variation due to protected attributes}

\allowdisplaybreaks

\begin{document}

\twocolumn[
\icmltitle{Debiasing representations by removing unwanted variation due to protected attributes
}



\icmlsetsymbol{equal}{*}

\begin{icmlauthorlist}
\icmlauthor{Amanda Bower}{equal,m}
\icmlauthor{Laura Niss}{equal,m}
\icmlauthor{Yuekai Sun}{equal,m}
\icmlauthor{Alexander Vargo}{equal,m}
\end{icmlauthorlist}

\icmlaffiliation{m}{University of Michigan, Michigan, USA}

\icmlcorrespondingauthor{}{}

\icmlkeywords{fairness, RUV, COMPAS, ICML}

\vskip 0.3in
]



\printAffiliationsAndNotice{\icmlEqualContribution} 

\begin{abstract}


We propose a regression-based approach to removing implicit biases in representations. On tasks where the protected attribute is observed, the method is statistically more efficient than known approaches. Further, we show that this approach leads to debiased representations that satisfy a 
first order approximation of conditional parity. Finally, we demonstrate the efficacy of the proposed approach by reducing racial bias in recidivism risk scores.
\end{abstract}
\section{Introduction}

In practice, the use of algorithms does not remove all human bias from decision making. There are numerous examples of algorithms with outcomes that are unfair to members of different protected classes (see e.g. \citet{angwin2016machine}, \citet{steel2010cutting}).
In recent years, there has been a flurry of work aimed at correcting this issue.  Starting with \cite{dwork2011fairness}, this work has generally fallen into four categories:
\begin{enumerate}
  \setlength{\itemsep}{1pt}
  \setlength{\parskip}{0pt}
  \setlength{\parsep}{0pt}
\item Mathematical or statistical definitions of fairness (e.g. \citet{frieder2016imposs}, \citet{ritov2017conditionalp}).
\item Algorithms that are modeled to ensure fairness (e.g. \citet{joseph2016bandits}).
\item Methods of preprocessing data in order to remove inherent bias so that algorithms trained on the debiased data will be fair (e.g. \citet{zemel2013learning}, \citet{feldman2015certifying}).
\item Methods of debiasing the outcomes of existing algorithms (a postprocessing step; e.g. \citet{hardt2016equality}).
\end{enumerate}

This work falls into the third category of preprocessing. We introduce a factor model prevalent in genetics applications to model the contributions of the protected and permissible attributes to the representation. By treating the variation that is present in the data due to protected attributes (e.g. race) as unwanted, we devise a method to remove this unwanted variation based on the factor model (and thus debias the data). We show that under certain idealized conditions, the debiased representation is conditionally uncorrelated with the protected attributes. In other words, it satisfies a first order approximation of conditional parity \cite{ritov2017conditionalp} in these cases.

\subsection{Motivating example}
We use ProPublica's COMPAS dataset and COMPAS risk recidivism scores as an example throughout. More information can be found from \cite{angwin2016machine} and the Practitioners Guide to COMPAS \footnote{http://www.northpointeinc.com/files/technical
\newline \_documents/FieldGuide2\_081412.pdf}. Much has been written questioning the fairness of these scores with respect to race, with concerns about the disparate false negative and false positive rates between African-Americans and Caucasians.

{\bf Notation:} We denote matrices by uppercase greek or Latin characters and vectors by lowercase characters. A (single) subscript on a matrix indexes its rows (unless otherwise stated). A random matrix $X\in\reals^{n\times d}$ is distributed according to a {\it matrix-variate normal} distribution with mean $M\in\reals^{n\times d}$, row covariance $\Sigma_r\in\reals^{n\times n}$, and column covariance $\Sigma_c\in\reals^{d\times d}$, which we denote by $X\sim\MN(M,\Sigma_r,\Sigma_c)$.

\section{Related Work}

The factor model that motivates the proposed approach (Equation \ref{eq:linearModel}) is widely used in genetics
 and was first introduced by \cite{Leek2008general} to represent wanted and unwanted variation in gene expression data. It was further exploited by \cite{gagnon2013removing} to develop the removing unwanted variation (RUV) family of methods.

RUV methods rely on knowledge of a set of control genes: genes whose variation in their expression levels are solely attributed to variation in $Z$, for example, genes unaffected by the treatments. Formally, a set of controls is a set of indices $\cI\subset[d]$ such that $B_{\cI} = 0$. Thus
\[
Y_{\cI} = XA_{\cI}^T + E_{\cI},
\]
where $Y_{\cI}$ and $E_{\cI}$ consist of subsets of the {\it columns} of $Y$ and $E$, which suggests estimating $A_{\cI}^T$ by linear regression. This is precisely the ``transpose'' of the method that we advocate. 

The task of debiasing features was studied by \cite{Lum2016statistical}, but there are a few differences between their approach and the proposed one. First, their approach creates new features that satisfy unconditional rather than conditional parity. Second, their approach entails estimating the conditional distributions of the features. This is hard in general, especially if the features are high-dimensional. 

\section{Adjusting for protected attributes}
\label{sec:debiasing}


Consider the widely adopted model for matrix-variate data:
\begin{equation}
\underset{(n\times d)}{Y} = \underset{(n\times k)}{X}\underset{(k\times d)}{A^T} + \underset{(n\times l)}{Z}\underset{(l\times d)}{B^T} + \underset{(n\times d)}{E}.
\label{eq:linearModel}
\end{equation}
The rows of $Y$ are representations, the rows of $X$ (resp.\ $Z$) are protected attributes (resp.\ permissible attributes) of the samples, and the rows of $E$ are error terms that represent idiosyncratic variation in the representations.
In this paper, we assume $k,l \ll d$, but this low-dimensionality is not requisite for the use of the algorithms presented here.

In practice, $Y$ is usually observed, $X$ is sometimes observed, and $Z$ is unobserved. For example, in \cite{Bolukbasi2016Man}, the representations are embeddings of words in the vocabulary, and the protected attribute is the gender bias of (the embeddings of) words. The rows of $Z$ are unobserved factor loadings that represent the ``good'' variation in the word embeddings. In analogy to the framework in \cite{frieder2016imposs}, the rows of $Z$ are points in the construct space while the rows of $Y$ are points in the observed space. We emphasize that like the construct space, $Z$ is unobserved.  

\begin{figure}
\vskip 0.2in
\begin{center}
\centerline{
  \tikz{
    \node[latent]                               (x) {$x$};
    \node[latent, below=of x, xshift=-1.2cm] (z) {$z$};
    \node[latent, below=of x, xshift=1.2cm]     (y) {$y$};
    \edge {x} {y,z}
    \edge {z} {y}
  }
}
\caption{The model \eqref{eq:linearModel} and \eqref{eq:confounding}. permissible attributes}
\label{fig:confounding}
\end{center}
\vskip -0.2in
\end{figure}

We highlight that we permit non-trivial correlation between the protected and permissible attributes. In other words, we allow the protected attribute to {\it confound} the relationship between the permissible attribute and the representation (see Figure \ref{fig:confounding} for a graphical representation of the dependencies between the rows of $Y$, $X$, and $Z$). This complicates the task of debiasing the representations. To keep things simple, we assume the regression of $Z$ on $X$ is linear:
\begin{equation}
\underset{(n\times l)}{Z} = \underset{(n\times k)}{X}\underset{(k\times l)}{\Gamma^T} + \underset{(n\times l)}{W}.
\label{eq:confounding}
\end{equation}
The rows of $W$ are error terms that represent variation in the permissible attributes not attributed to variation in the protected attributes. 
We specify the distributions of $X$, $E$, and $W$, in Sections \ref{sec:protectedAttributeLatent} and \ref{sec:protectedAttributeObs}.

{\it Our goal is to obtain debiased representations $Y_{\db}$ such that the debiased representations are uncorrelated with the protected attributes conditioned on the permissible attributes:}
\begin{equation}
\Cov\bigl[[Y_{\db}]_i,x_i\mid z_i\bigr] = 0.
\label{eq:conditionallyUncorrelated}
\end{equation}
This is implied by {\it conditional parity}: $[Y_{\db}]_i\perp x_i\mid z_i$, and we consider \eqref{eq:conditionallyUncorrelated} as a first-order approximation of conditional parity. An ideal debiased representation is the variation in the representation attributed to the permissible attributes $ZB^T$, but this is typically unobservable in practice. 

\subsubsection{COMPAS example}\label{sec:compas_ex_1}
Under this model, each row of $Y$ corresponds to a person's data for recidivism prediction. In our experiments, this includes age, juvenile and adult felony and misdemeanor counts, and whether the offense was a misdemeanor or felony. In this case, $X$ is a vector, and each component indicates the  person's race. We restrict to Caucasians and African-Americans in our experiments. A persons's true propensity to choose to commit a crime, $Z$, is unknown.

\subsection{Homogeneous subgroups}

The proposed approach relies crucially on knowledge of homogeneous subgroups: groups of samples in which the variation in their representations is mostly attributed to variation in their protected attributes. Formally, we presume knowledge of sets of indices $\cI_1,\dots,\cI_G\subset[n]$ such that $H_gZ_{\cI_g} \approx 0$, where $H_g =I_{\abs{\cI_g}} - \frac{1}{\abs{\cI_g}}\ones_{\abs{\cI_g}}\ones_{\abs{\cI_g}}^T$
is the centering matrix, for any $g\in[G]$. In other words,
\[
H_gY_{\cI_g} \approx H_gX_{\cI_g}A^T + H_gE.
\]
Ideally, $H_gZ_{\cI_g}$ exactly vanishes. This ideal situation arises when the samples in the $g$-th group share permissible attributes: $Z_{\cI_g} = \ones_{\abs{\cI_g}}z_g^T$
for some $z_g\in\reals^l$.

Intuitively, homogeneous subgroups are groups of samples in which we expect a machine learning algorithm that only discriminates by the permissible attributes to treat similarly. For example, in \cite{Bolukbasi2016Man}, the homogeneous subgroups are pairs of words that differ only in their gender bias: ({\it waiter}, {\it waitress}), ({\it king}, {\it queen}).
\subsubsection{COMPAS example}
In Section \ref{sec:experiments}, we take the homogeneous groups to be people who either did not recidivate within two years or people who did recidivate within two years and were charged with the same degree of felony or misdemeanor. Although $Z$ is unknown, we expect subjects who go on to commit similar crimes or those who do not recidivate to have similar $Z$ regardless of race. We emphasize that the homogeneous subgroups are not defined by having similar attributes in $Y$. 

\subsection{Adjustment when the protected attribute is unobserved}
\label{sec:protectedAttributeLatent}

We now show that the approach proposed by \citet{Bolukbasi2016Man} produces debiased representations that satisfy \eqref{eq:conditionallyUncorrelated}.
When the protected attributes are not observed, it is generally not possible to attribute variation in the representations to variation in the protected and permissible attributes.
Thus, \citet{Bolukbasi2016Man} settle on removing the variation in the representations in the subspace spanned by the protected attributes. In other words, we debias the representations by projecting them onto the orthocomplement of $\cR(A)$. 


Formally, let $Q_g\in\reals^{\abs{\cI_g}\times(\abs{\cI_g} - 1)}$ be a subunitary matrix such that $\cR(Q_g)$ coincides with $\cR(H_g)$. Under \eqref{eq:distributionalAssumptions},

\begin{equation*}
\begin{aligned}
&Q_g^TY_{\cI_g} \approx Q_g^TX_{\cI_g} + Q_g^TE,
\end{aligned}
\end{equation*}

which implies $\Cov\bigl[Q_g^TY_{\cI_g}\bigr] \approx \Sigma_E + AA^T$. This is a factor model, which allows us to consistently estimate $A$ by factor analysis under mild conditions. We impose classical sufficient conditions for identifiability of $A$ \cite{Anderson1956Statistical}:
\begin{enumerate}
  \setlength{\itemsep}{1pt}
  \setlength{\parskip}{0pt}
  \setlength{\parsep}{0pt}
\item Let $A_{-i}$ be the $(d-1)\times k$ submatrix of $A$ consisting of all but the $i$-th row of $A$. For any $i\in[n]$, there are two disjoint submatrices of $A_{-i}$ of rank $k$.
\item $A^T\Sigma_E^{-1}A$ is diagonal, and the diagonal entries are distinct, positive, and arranged in decreasing order.
\end{enumerate}

We remark that the additional assumptions we imposed in this section are a tad stronger than necessary: the assumptions actually imply identifiability of $A$, but we only wish to estimate $\cR(A)$.

In light of the preceding development, here is a natural approach to adjustment when the protected attribute is unobserved:
\begin{enumerate}
  \setlength{\itemsep}{1pt}
  \setlength{\parskip}{0pt}
  \setlength{\parsep}{0pt}
\item estimate $A$ by factor analysis: \\
$\argmin\{{\textstyle\frac12\sum_{g=1}^G\|H_gY_{\cI_g} - XA\|_F^2}\}$;
\item debias $Y$ by projection onto $\cR(A)^\perp$: \\
$Y_{\db} = Y(I - P_{\cR(A)})$,
\end{enumerate}

which gives
\[
\Cov[[Y_{\db}]_i, x_i|z_i] = \Cov[P_{\cR(A)^{\bot}}(Bz_i + e_i)|z_i] = 0.
\]

Note that when $B \subset \cR(A)$ the debiased representations will be non-informative because they only contain noise.

\subsection{Adjustment if the protected attribute is observed}
\label{sec:protectedAttributeObs}

If the protected attribute is observed, it is straightfoward to debias the representations. The main challenge here is estimating $A$. Once we have a good estimator $\hA$, we debias the representations by subtracting $X\hA^T$. We summarize the
approach in Algorithm \ref{alg:protectedAttributeObs}.

\begin{algorithm}
   \caption{Adjustment if the protected attr. is observed}
   \label{alg:protectedAttributeObs}
\begin{algorithmic}
   \STATE {\bf Input:} representations $Y\in\reals^{n\times d}$, protected attributes $X\in\reals^{n\times k}$ and groups $\cI_1,\dots,\cI_G\subset[n]$
   \STATE {\bf Estimate $A$ by regression:}
   \[
   \hA^T \in \argmin\{{\textstyle\frac12\sum_{g=1}^G\|Y_g - X_gA^T\|_F^2}\},
   \]
   where $Y_g = Y_{\cI_g} - \ones_{\abs{\cI_g}}(\frac{1}{\abs{\cI_g}}\ones_{\abs{\cI_g}}^TY_{\cI_g})$ and $X_g$ is defined similarly.
   \STATE {\bf Debias $Y$:} subtract the variation in $Y$ attributed to $X$ from $Y$: $Y_{\db} = Y - X\hA^T$.
\end{algorithmic}
\end{algorithm}

To study the properties of Algorithm \ref{alg:protectedAttributeObs}, we impose the following assumptions on the distributions of $X$, $E$, and $W$:
\begin{equation}
\begin{aligned}
X \sim \MN(0,I_n,\Sigma_x), \\
E\mid(X,Z) \sim \MN(0,I_n,\Sigma_{\eps}).
\end{aligned}
\label{eq:distributionalAssumptions}
\end{equation}

\begin{proposition}
Let $Z_g$ and $E_g$ be defined similarly as $Y_g$ and $X_g$. Under conditions \eqref{eq:linearModel}, \eqref{eq:confounding}, and \eqref{eq:distributionalAssumptions},
\[
\begin{aligned}
&\hA^T - A^T\mid(X,Z) \sim \MN({\textstyle{T\sum_{g=1}^GX_g^TZ_gB^T,   T, \Sigma_\eps}})
\end{aligned}
\]
where $T = (\sum_{g=1}^GX_g^TX_g)^\dagger$.
\end{proposition}


The (conditional) bias in the OLS estimator of $A$ depends on the similarity of the permissible attributes in homogeneous subgroups. If $Z_g = 0$ for all $g\in G$,
then $\hA$ is a (conditionally) unbiased estimator of $A$.

\begin{proposition}
Under conditions \eqref{eq:linearModel}, \eqref{eq:confounding}, and \eqref{eq:distributionalAssumptions}, we have
\[
\Cov[y_i - \hA x_i,x_i\mid z_i] = -B\Cov[\tilde{Z}\tilde{X}(\tilde{X}^T\tilde{X})^\dagger x_i,x_i|z_i].
\]
\end{proposition}

We see that if $\hat{A} = A$ or $\tilde{Z} = 0$ then the debiased $y_i$ is uncorrelated with the protected attributes $x_i$. 
\section{Experiments: Debiased representations for recidivism risk scores}\label{sec:experiments}

\begin{figure}[]
    \centering
    \includegraphics[width=0.5\textwidth, clip=true, trim = {0, 1.2cm, 0, 0}]{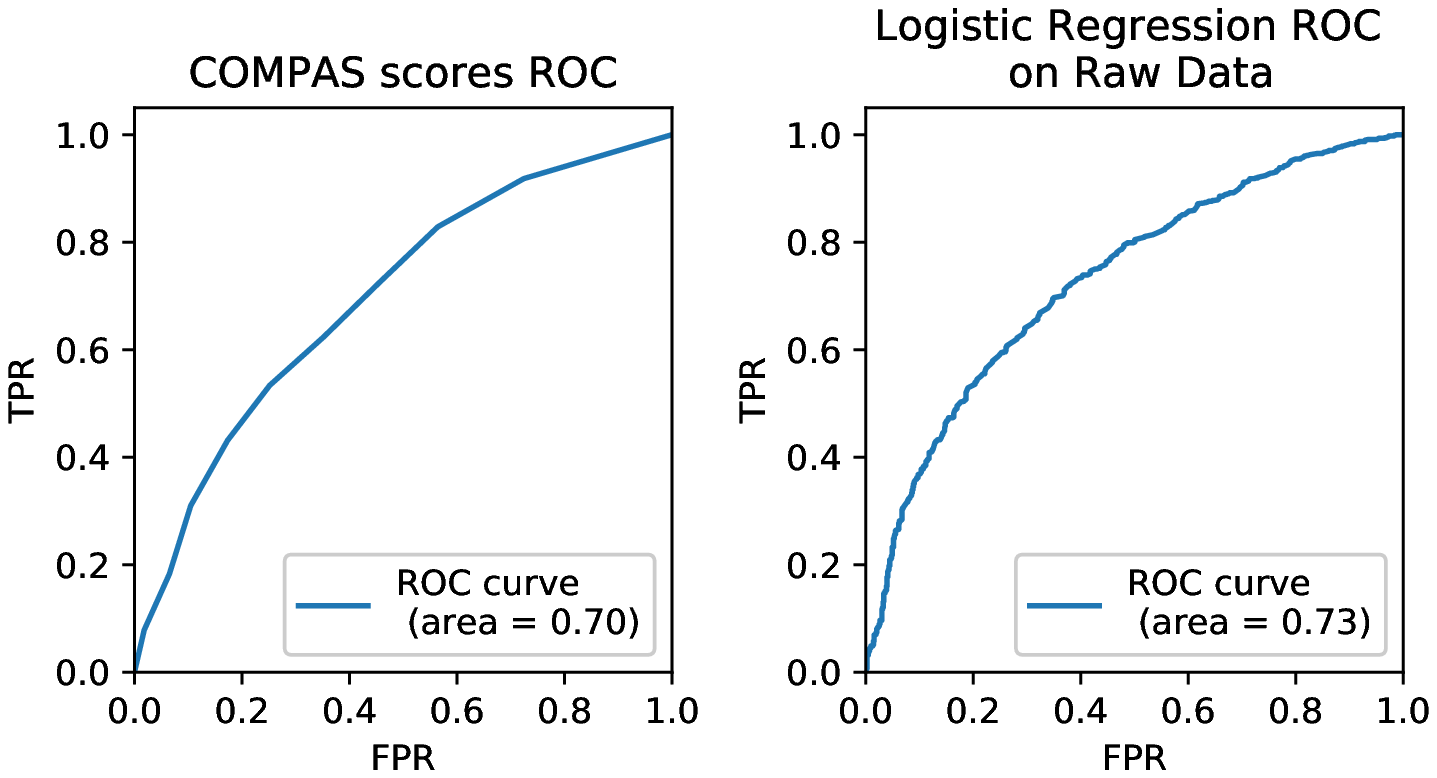}
    \caption{}
    \label{fig:biased_ROC}
\end{figure}

We empirically demonstrate the efficacy of Algorithm \ref{alg:protectedAttributeObs} for reducing racial bias in  recidivism risk scores based on data ProPublica \footnote{https://github.com/propublica/compas-analysis/blob/master/compas-scores-two-years.csv} used in their investigation of COMPAS scores. 
We fit our own models to the raw and debiased data. Although simple, the scores output by our model perform comparably to the proprietary COMPAS scores (see Figure \ref{fig:biased_ROC} and Table \ref{table:acc}).

In particular, we show that logistic regression (LR) trained on debiased data obtained from Algorithm \ref{alg:protectedAttributeObs} reduces the magnitude of the difference in the false positive rates (FPR) and false negative rates (FNR) between Caucasians and African-Americans (AA) compared to LR trained on raw, potentially biased data. This ``fairer'' outcome is achieved with a relatively small impact on the percentage of correct predictions. The variables in our LR model are discussed in Section \ref{sec:compas_ex_1}.

\begin{figure}[]
    \centering
    \includegraphics[width=0.5\textwidth]{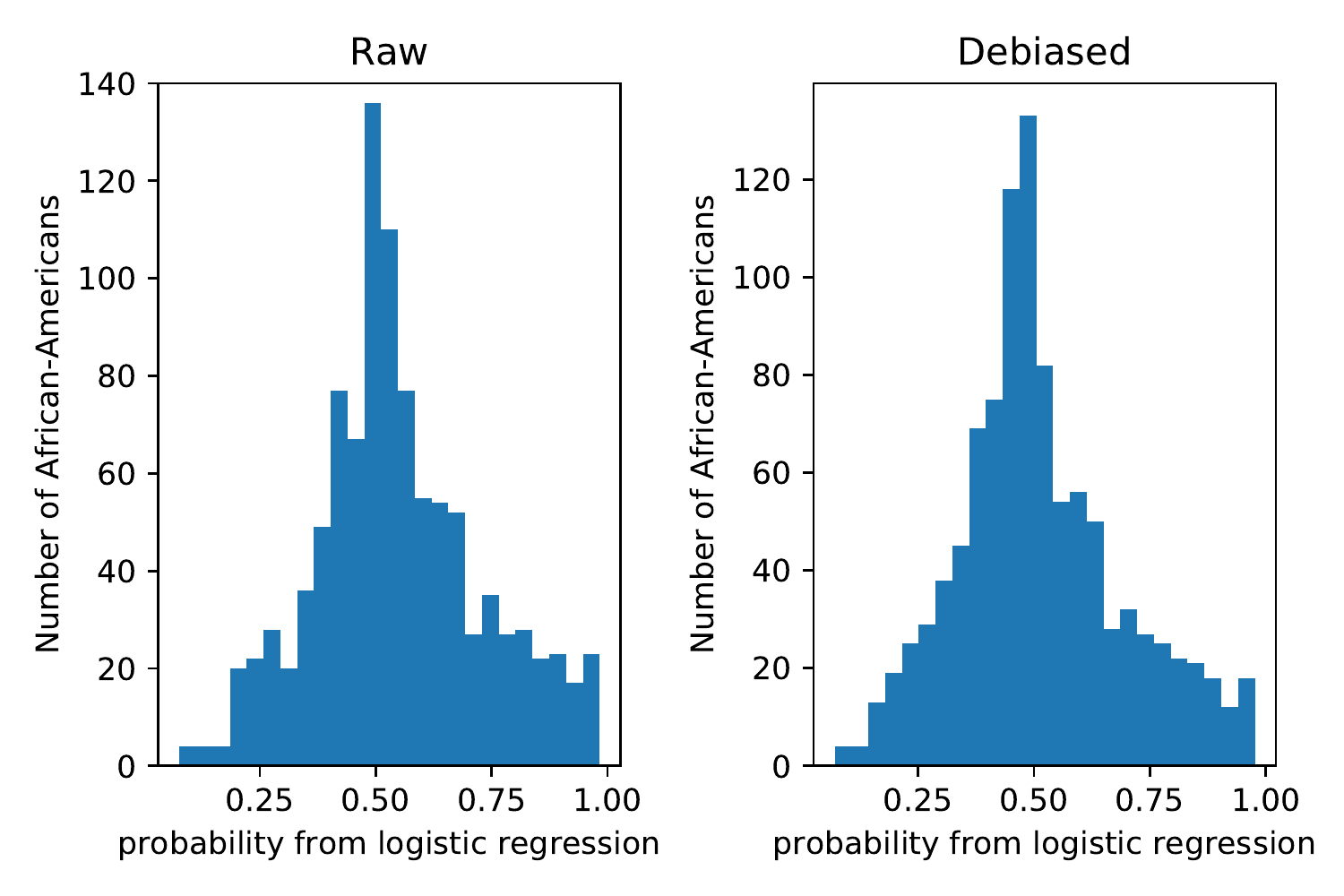}    \caption{Distribution of recidivism probabilities from raw and debiased representations for African-Americans. }
    \label{fig:histo.pdf}
\end{figure}

We split our data into three pieces: a training set used to estimate $A$ from Equation \eqref{eq:linearModel}, and a train and test set to evaluate the performance of the learned model. Figure \ref{fig:histo.pdf} shows the distribution of the probabilities of recidivism for African-Americans according to a logistic model based on the raw and debiased representations. The distribution of the probabilities from the raw representation is skewed to the right. In particular, the right tail of the distribution of probabilities from the raw representation is noticeably heavier than that from the debiased representation.

The ROC curve of the LR model trained on raw data is similar to the ROC curve of COMPAS scores validating the choice of LR as a proxy for COMPAS scores. See Figure \ref{fig:biased_ROC}. 
For the remaining discussion, we average all results over 30 splits of the data into train and test sets. The average accuracy (the percentage of correct predictions) is 65\% for COMPAS. The accuracy for LR trained on raw data and debiased data is comparable, again validating our proxy and justifying the slight loss in accuracy after debiasing in pursuit of fairer outcomes. See Table \ref{table:acc}.

 Table \ref{table:80_quantile} and Table \ref{table:50_quantile} show the average FPR and FNR for the LR model before and after debiasing. The two tables differ only in the threshold used to declare someone at risk for recidivism based on his or her logistic score; we choose to examine the 50th and 80th quantiles of LR scores since Northpointe specifies that COMPAS scores above the of 50th (respectively 80th) quantile are said to indicate a ``Medium"  (respectively ``High") risk of recidivism. In Table \ref{table:80_quantile}, we see that there is no difference in FPR after debiasing. The difference in FNR between the races goes from nearly 20\% before debiasing to 4\% after debiasing. In Table \ref{table:50_quantile}, we see FNR are nearly equalized, whereas the magnitude of the difference of FPR between both race groups is improved. However, now Caucasians suffer from disparate impact of FPR instead of African-Americans. 

\begin{table}[h!]
\centering
\resizebox{\columnwidth}{!}{\begin{tabular}{l|l|l|l|l|}
\cline{2-5}
                                 & \multicolumn{2}{c|}{LR raw} & \multicolumn{2}{c|}{LR debiased} \\ \cline{2-5} 
                                 & FPR (SE)     & FNR (SE)     & FPR (SE)       & FNR (SE)        \\ \hline
\multicolumn{1}{|l|}{Population} & 0.8 (0.01)    & 0.68 (0.01)   & 0.9 (0.01)      & 0.69 (0.01)      \\ \hline
\multicolumn{1}{|l|}{Caucasian}  & 0.05 (0.01)    & 0.81 (0.02)   & 0.9 (0.02)      & 0.72 (0.03)      \\ \hline
\multicolumn{1}{|l|}{AA}         & 0.11 (0.01)   & 0.62 (0.01)   & 0.9 (0.01)      & 0.68 (0.02)      \\ \hline
\end{tabular}}
\caption{
 Average proportion FPR and FNR with standard errors (SE) based on the 80th quantile of LR scores.
}
\label{table:80_quantile}
\end{table}

\begin{table}[h!]
\centering

\resizebox{\columnwidth}{!}{\begin{tabular}{l|l|l|l|l|}
\cline{2-5}
                                 & \multicolumn{2}{c|}{LR raw} & \multicolumn{2}{c|}{LR debiased} \\ \cline{2-5} 
                                 & FPR (SE)     & FNR (SE)     & FPR (SE)        & FNR (SE)       \\ \hline
\multicolumn{1}{|l|}{Population} & 0.32 (0.01)   & 0.32 (0.01)   & 0.4 (0.01)      & 0.34 (0.01)     \\ \hline
\multicolumn{1}{|l|}{Caucasian}  & 0.22 (0.02)   & 0.5 (0.02)   & 0.42 (0.03)      & 0.31 (0.02)     \\ \hline
\multicolumn{1}{|l|}{AA}         & 0.4 (0.02)   & 0.23 (0.01)   & 0.27 (0.02)      & 0.35 (0.02)     \\ \hline
\end{tabular}}
\caption{
 Average proportion FPR and FNR with standard errors (SE) based on the 50th quantile of LR scores.
}
\label{table:50_quantile}
\end{table}


\begin{table}[h!]
\centering
\resizebox{\columnwidth}{!}{\begin{tabular}{l|l|l|l|}
\cline{2-4}
\textbf{Accuracy}                 & LR Raw (SE) & LR Debiased (SE) & COMPAS (SE) \\ \hline
\multicolumn{1}{|l|}{50 quantile} & 0.67 (.011)  & 0.65   (.01)     & 0.65 (.008) \\ \hline
\multicolumn{1}{|l|}{80 quantile} & 0.61 (.01)   & 0.60  (.01)      & 0.61 (.01)  \\ \hline
\end{tabular}}
\caption{
 Proportion of correct predictions (with standard errors) by logistic regression and thresholding COMPAS scores
}
\label{table:acc}
\end{table}

\section{Summary and discussion}

We study a factor model of representations that explicitly models the contributions of the protected and permissible attributes. Based on the model, we propose an approach to debias the representations. We show that under certain conditions, we can guarantee 
first order
conditional parity for the debiased representations. 


\bibliographystyle{icml2018}
\Urlmuskip=0mu plus 1mu\relax
\bibliography{bibliography,yuekai}

\end{document}